\documentclass{article}
\usepackage{graphicx} 
\usepackage{xcolor}
\usepackage{hyperref}
\usepackage{authblk}
\usepackage{listings}
\usepackage{multicol}
\hypersetup{
    colorlinks=true,
    linkcolor=blue,
    citecolor=blue,     
    urlcolor=blue,
    }
    
\title{Accelerating the Dutch Atmospheric Large-Eddy Simulation (DALES) model with OpenACC}

\author[1]{Lucas Esclapez}
\author[1]{Laurent Soucasse}
\author[2]{Caspar Jungbacker}
\author[2]{Fredrik Jansson}
\author[2]{Stephan R. de Roode}
\author[3]{Pedro Costa}
\author[1]{Gijs van den Oord}
\author[1]{Alessio Sclocco}

\affil[1]{Netherlands eScience Center, Amsterdam, The Netherlands}
\affil[2]{Dept of Geoscience \& Remote Sensing, Delft University of Technology, Delft, The Netherlands}
\affil[3]{Process \& Energy Department, Delft University of Technology, Delft, The Netherlands}

\begin{document}

\maketitle

\begin{abstract}
This paper presents the GPU porting through OpenACC directives of the Dutch Atmospheric Large-Eddy Simulation (DALES) application, a high-resolution atmospheric model. The code is written in Fortran~90 and features parallel (distributed) execution through spatial domain decomposition. We assess the performance of the GPU offloading, 
comparing the time-to-solution on regular and accelerated HPC nodes. 
A weak scaling analysis is conducted and portability across NVIDIA A100 and H100 hardware 
is discussed. Finally, we show how  targeted kernels can benefit from further optimization with Kernel Tuner, a GPU kernels auto-tuning package.
\end{abstract}

\section{Introduction}
\label{sec:intro}

Atmospheric models are among the most challenging physical models to solve numerically. Model complexity and associated computational costs come from the chaotic nature of the airflow dynamics and its coupling with various transport processes such as precipitation, chemical reactions, solar and thermal infrared radiation and ocean/land interactions. The accuracy and reliability of the simulations strongly depend on the spatial and temporal resolutions with which the model is discretized and, in return, on the efficient use of significant computational resources.

Atmospheric numerical models naturally rely on parallel implementation on powerful supercomputers. The spatial discretization employed in atmospheric models is well suited for domain decomposition, with independent mesh partitions distributed on compute clusters using MPI. The last decade has seen the massive use of GPU accelerators in high performance computing (HPC), and GPUs represent 3/4 of the computational power of the machines in the Top500 list today~\cite{TOP500}. The climate community has thus explored different ways to efficiently use GPUs.

The community early work focused on converting selected computationally expensive kernels to CUDA \cite{Michakales2008}, leading to up to 20x speedup for given compute kernels and 1.3x speedup at the solver level (on a single CPU versus single GPU basis). Subsequent efforts aimed at accelerating entire solvers and led to the development of solvers such as NIM-CUDA \cite{Govett2010}, GALES \cite{Schalkwijk2012} or MicroHH \cite{vanHeerwaarden2017}, the later two introducing a C++/CUDA re-write of a previous Fortran code base. However, such drastic approaches require significant code development investments and disruptive changes for the users. In recent years, directive-based acceleration using OpenACC has been successfully applied to weather and climate Fortran code bases \cite{Lapillonne2014, Norman2015, Giorgetta2022}, striking a compromise between implementation efforts and performance on GPU.

This paper addresses the GPU porting of a high-resolution atmospheric model called the Dutch Atmospheric Large-Eddy Simulation (DALES) model through OpenACC directives. We first present the DALES implementation and our acceleration strategy in Sec.~\ref{sec:DALES}. The simulation experiments we perform to assess the offloading performance are described in Sec.~\ref{sec:exp}. We then discuss the offloading results in Sec.~\ref{sec:perf}, obtained on a single-node basis and different hardware, namely NVIDIA A100 and NVIDIA H100 (Snellius supercomputer). 
This section also covers the multi-node performance and a weak scaling analysis. Finally, we address how Kernel Tuner~\cite{vanwerkhoven19}, an auto-tuning GPU application, can further improve the performance in Sec.~\ref{sec:opt}, and we conclude in Sec.~\ref{sec:conclu}.

\section{DALES framework}
\label{sec:DALES}

\subsection{Modelling}

DALES is an atmospheric large-eddy simulation code, used to model atmospheric processes such as convection, clouds, and precipitation. It is generally run with a horizontal resolution of 10 to 200~m and a vertical resolution of 10 to 50~m, meaning that it can explicitly simulate clouds and convective updrafts. The model combines a large-eddy simulation of fluid dynamics with models of physical processes in the atmosphere. The dynamics is solved for incompressible air, in the anelastic approximation~\cite{Boing2012}.

The fluid dynamics part contains advection of momentum and scalars in finite difference formulation. Several advection schemes are available, see Ref.~\cite{heus10}. Time integration is performed by an explicit third order Runge-Kutta scheme.
Since the flow is assumed incompressible, a Poisson equation for the pressure field is solved at each time step to ensure mass conservation.
As the spatial discretization along horizontal directions uses a regular grid spacing, a FFT-based approach for the finite-difference Poisson equation is used in the lateral directions~\cite{costa18}.
Subgrid turbulence, i.e. diffusion caused by turbulent eddies smaller than the grid size, is modelled with the Deardorff scheme~\cite{Deardorf1980}. A prognostic model variable for the subgrid turbulent kinetic energy is used to estimate diffusion coefficients for momentum and other transported quantities~\cite{de2017diagnosis}. Such a \emph{subgrid-scale model} is almost what defines a large-eddy simulation: the model resolves the ``large" turbulent eddies and parameterizes effects of the smaller ones.

The physical processes include:
\begin{itemize}
    \item buoyancy of air dependent on temperature and moisture content;
    \item condensation of water vapor, forming cloud droplets and releasing latent heat, hereafter referred to as ``thermodynamics"; 
    \item ``cloud microphysics", i.e. the conversion of cloud droplets to rain droplets, and other processes related to rain or snow;
    \item radiative transfer, solved using the independent column approximation within both the infrared range (longwave) and visible solar light range (shortwave). There are several schemes for simulating the radiative transfer. One of them is RTE-RRTMGP~\cite{pincus19}, which we focus on here since it supports offloading to GPUs;
    \item atmospheric chemistry, modelling chemical reaction between chemical species in the atmosphere (optional and not considered in this article).
\end{itemize}

\subsection{Implementation}
DALES is written in Fortran~90 and employs an MPI-based domain decomposition approach for parallelism. The code base consists of about 75k lines of codes, organised in physics, numerics and utility modules. The basic data structures, stored in data modules, are Fortran multi-dimensional arrays which are allocated only once upon initialization since the computational grid is fixed. 
3D-arrays are indexed \emph{ijk}, meaning the  \emph{k} indices associated with the vertical direction are not addressed contiguously.
The domain decomposition partitions the computational grid into $z$-pencils (in the direction normal to the ground), generating  Cartesian sub-domains whose size in the horizontal directions is inversely proportional to the number of MPI ranks.
On each partition, halo cells are added in the horizontal directions to ensure data consistency across MPI ranks using non-blocking send/receive exchanges. The number of halo cells varies from 1 to 3 depending on the advection scheme stencil size. DALES does not feature shared-memory parallelization (e.g. using OpenMP).

DALES includes a bundled netlib fast Fourier Transform (FFT) module to ensure portability of the Poisson solver, but also relies on external libraries such as FFTW for increased efficiency and flexibility.
Finally, DALES employs NetCDF for most of its input/output operations except checkpointing which uses plain binary format.


\subsection{Acceleration}
\label{ssec:accel}

Owing to the size of the code base and the long duration of the tuning and validation of the model parameters, a complete overhaul of the solver in a more modern and GPU-friendly programming language (e.g. C++) is not an option. Additionally, DALES is used by researchers and students alike (see Ref.~\cite{frassoni2018building} for instance), with the later group often having a limited computer science background thus making Fortran's imperative programming style a good choice. OpenACC enables to leverage the high throughput of GPUs with limited disruption to the code base and has been successfully applied to several fluid and weather solvers over the past few years \cite{Xia2015,Norman2015,Giorgetta2022,Costa2021}. The following provides an overview of the main aspects of the DALES acceleration.

\subsubsection{Data management}
Data containers in DALES are distributed across the physics, numerics and utility Fortran modules, each module managing the dynamic memory allocation and deallocation of the multi-dimensional arrays necessary for its computations. In early GPUs (e.g. NVIDIA Kepler and older), the limited available device memory required careful management of the data loaded in the device memory for each kernel, and data transfers to and from the device was found to be a significant bottleneck software developers had to overcome in order to obtain good performance on GPUs. Leveraging the larger memory available on modern GPUs, the entire memory of the CPU is mirrored on the device in DALES, with each module using the \verb|enter data copyin| 
clause once after initialization of the data on the CPU, and clearing the device memory with \verb|exit data delete| clause before deallocating the arrays on the CPU. Thus, data is always assumed to be already present on the device when launching compute kernels and data copy from the device to the host memory only happens when performing diagnostics or checkpointing.

\subsubsection{Overall porting strategy}
DALES typical compute subroutines fall into one of the following categories:
\begin{itemize}
    \item nested independent \emph{ijk} or \emph{ijkn} loops, traversing the three spatial dimensions and possibly a fourth dimension in the scalar index space.
    \item similarly nested loops but carrying a dependency over the $k$ index corresponding to the vertical direction (normal to the ground).
\end{itemize}
The sizes of the \emph{ijk} loops are case dependent and can be found in a wide range of values when considering the typical use case of DALES. As such, a premature optimization of the OpenACC pragmas for a given case size is not considered. Rather, our default approach consists in using the \verb|collapse| clause on as many tightly nested loops as possible in order to expose a maximum of parallelism to the GPU while remaining portable, with the \emph{stride-1} inner loop index, as follows:

\begin{lstlisting}[language=Fortran, caption=Typical DALES triply-nested Fortran loop]
!$acc parallel loop collapse(3)
do k = 1, kmax
  do j = 1, jmax
    do i = 1, imax
      phi(i,j,k) = ...
    enddo
  enddo
enddo
\end{lstlisting}

Alternate OpenACC clauses, including changing the \verb|parallel loop| default parameters and caching will be discussed in Sec.~\ref{sec:opt}. Most of DALES compute kernels are concise, such that register pressure is not a limiting factor except in a few specific cases that will be discussed Sec.~\ref{sec:opt}. DALES \emph{ijk} loops often include an \verb|if| statement for the surface layer ($k=1$). To avoid branching within the device code and the resulting thread divergence, the $k=1$ case is launched in a separate kernel, using the \verb|async| clause to hide the launch latency. For \emph{ijk} loops carrying a $k$ dependency, atomic operations are used when the dependency is strictly additive (e.g. in the cloud sedimentation kernels, where the cloud droplets are transferred between adjacent $k$ levels) or a temporary array is introduced.

When possible, consecutive independent kernels are fused but DALES often intertwine compute kernels and statistics sampling in order to extract diagnostics able to provide insights into the complex physics at play. In this later case, \verb|async| clauses in separate queues allows to hide launch latency while keeping the kernels in separate subroutines for readability and convenience.

\subsubsection{Refactoring}

In order to enable collapsing as many loops as available, local refactoring of several compute kernels was necessary to displace $k$-only dependent computations into the innermost loop, at the small price of repeated calculations. More extensive refactoring was needed in microphysics and thermodynamics, where differences in the local state can introduce significant thread divergence. Two examples of such refactoring are:
\begin{itemize}
    \item All the rain microphysics computations in DALES are masked to take advantage of the observation that rain droplets are generally present in a limited range of altitude ($k$-levels) and clustered in "showers" in the spanwise directions. The actual range of $k$ in microphysics \emph{ijk} loops is based on a scan of the mask, initially relying on $k$-only loop combined with Fortran-array syntax and an \verb|exit| statement on CPU. This approach performs poorly on GPUs, where a reduction across a fully collapsed \emph{ijk} loops was used. In this case, separate versions of the code are needed since the GPUs approach proved more expensive on CPU.
    \item DALES uses a saturation adjustment scheme in the thermodynamics routines to diagnose cloud liquid water content. To adjust the liquid water potential temperature accordingly, a Newton-Raphson algorithm is needed. The local conditions in each computational cell can lead to a varying number of Newton iterations needed to reach the desired accuracy, introducing thread divergence. Replacing the iterative algorithm by explicitly unrolling a fixed number of iterations deemed sufficient for accuracy alleviated the divergence issue and resulted in a more efficient kernel on both CPU and GPU.
\end{itemize}

\subsubsection{External libraries} \label{sec:acc_librairies}
Alongside this acceleration effort, DALES relies on external libraries that already features GPU offloading.

Radiative transfer calculations are performed with the RTE-RRTMGP libraries~\cite{pincus19}. The library is made of two parts, both implemented in Fortran and featuring GPU porting through OpenACC directives. The first part (RRTMGP) computes the radiative properties of molecules and clouds from the input state of the atmosphere by interpolating tabulated data. The second part (RTE) solves the radiation problem and output radiative fluxes accross the spatial domain. Each part has to be performed independently for each vertical column and each spectral wavelength, which makes GPU acceleration efficient. However, it is in practice not possible to run all column calculations simultaneously because of memory constraints. The total number of columns is thus divided into several batches. It is important to maximize the number of columns per batch (or minimize the number of batch) as multiple calls to RRTMGP and RTE libraries induce extra computational costs associated with kernel launch overhead. 

In order to perform the 3D FFT required by the Poisson solver on the GPU, we rely on the cuFFT library. 
Fourier transforms are performed one direction of space after another. Because Fourier transforms are non local, it needs to access all the field data along the direction performed, across the domain partition. This requires reorganizing the 3D field memory layout by performing a transposition, which is communication intensive with all-to-all communications. The same applies for the inverse Fourier transform, once the Poisson equation for the Fourier modes is solved. Finally, we should note that cuFFT does not support FFTW's ``half-complex'' numbering format used in DALES for in-place real-to-complex transforms of size $N$. To circumvent this, we adopted the approach used in the CaNS~\cite{Costa2021} and employed a minimal pre-/post-processing step of cuFFT's input/output, along with a re-ordering of the modified wavenumbers.


\section{Simulation experiments}
\label{sec:exp}

\subsection{Reference test case}
\begin{figure}
    \centering
    \includegraphics[width=\linewidth]{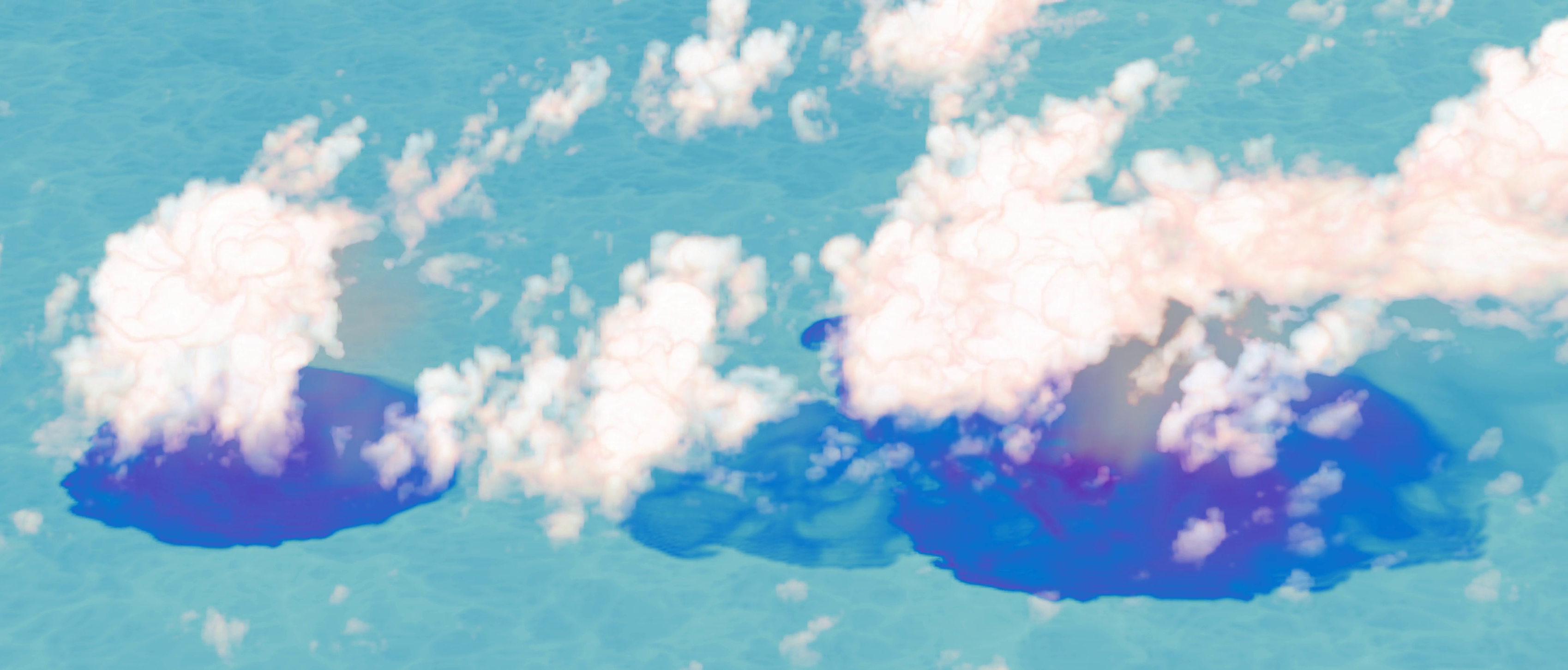}
    \caption{3D visualization of the clouds in one Cloud Botany simulation. Clouds are shown in white and rain is shown in gray. The air temperature near the surface is shown in blue, dark blue areas being cold pools associated with rain.}
    \label{fig:botany-scene}
\end{figure}

As a test case for the GPU porting project we chose one simulation from the \emph{Cloud Botany} ensemble of simulations. \emph{Cloud Botany}~\cite{Jansson2023} is a set of 103 DALES simulations of shallow cumulus clouds over the subtropical ocean, exploring how the clouds and their mesoscale organization behave under different physical conditions, for example changing sea surface temperature or wind speed. The simulated domains were 150~km wide in both horizontal directions, large enough to permit organization on a larger scale (called mesoscale in this context, for scales between single clouds and the scale of whole weather systems). One objective of the ensemble was to evaluate how mesoscale cloud organization affects the cloud-climate feedback~\cite{Janssens2024preprint}, i.e. to assess whether these clouds in a warmer climate will reflect more or less sunlight back into space. Figure~\ref{fig:botany-scene} shows a typical 3D visualization  of clouds and rain that can be extracted from the \emph{Cloud Botany} simulations.

This reference test case is useful to accelerate with GPUs, because: (i) it utilizes the core of the model and the optional physical models for solar radiative transfer and rain; (ii) it is a case where it is scientifically interesting to make the simulation domain as large as computationally feasible: a larger domain gives more space for mesoscale organization.

In the following simulation experiments, the grid size in the horizontal directions is fixed to 100~m but the number of points is changed between cases in order to address different problem sizes (in the range 50-100~km). However, the vertical domain size and spatial discretisation remain the same for all tests, namely a height of 7~km discretised using 175 points. As a first step, the model is integrated in time for about 24 to 36~h (model time) in order to reach a statistically steady state. 
All simulations will be run from this state for a model integration time of 10~min to make representative performance assessments.
We made sure this statistically steady state falls at day time so that it features shortwave radiation. 

\subsection{Verification}

\begin{figure*}[!ht]
    \centering
    \includegraphics[width=0.98\linewidth]{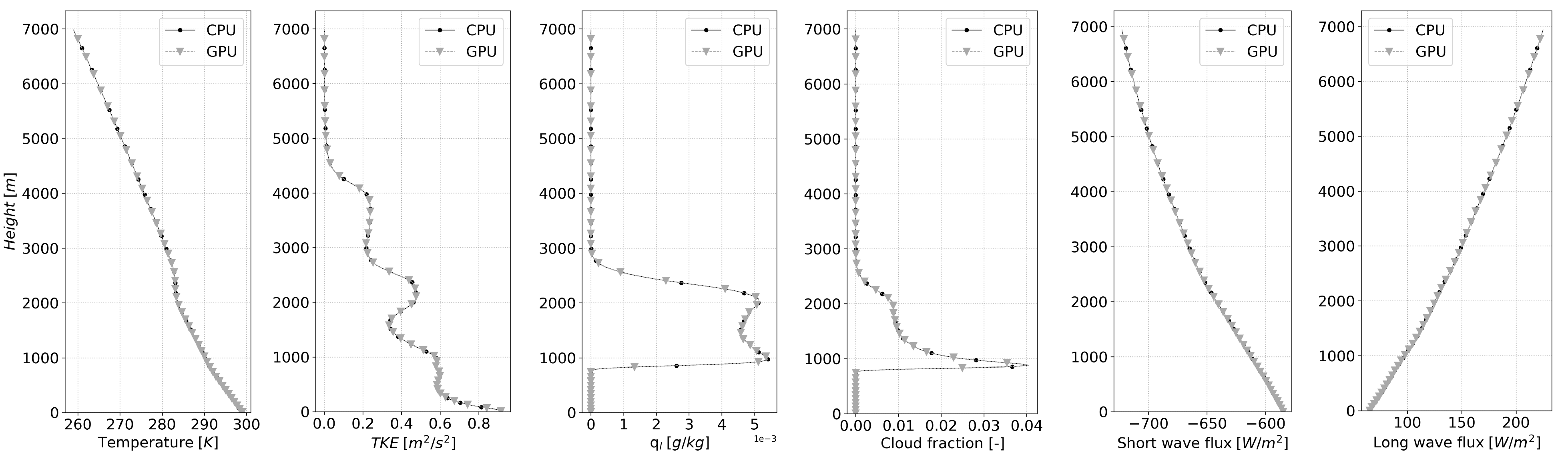}
    \caption{Vertical profiles of time and planar-averaged fields: temperature, resolved turbulent kinetic energy (TKE), liquid water specific humidity $q_l$, cloud fraction, shortwave and longwave net radiative fluxes (counted positively from Earth to space). Plot for both GPU and CPU implementation. 
    }
    \label{fig:cpuvsgpu_verification}
\end{figure*}

To verify the implementation of the OpenACC directives we checked correctness of the model output between CPU and GPU implementations. Due to the chaotic nature of the system, round-off errors quickly lead to different simulation trajectories. We will thus base the comparison on space and time-averaged statistics, which should remain close. Starting from the same statistically steady state, the simulation is pursued using  either a single CPU (AMD Rome 7H12) or a single GPU (NVIDIA A100) for 30~mn model time and a spatial resolution of about 500$\times$500 points in the horizontal directions, accumulating statistics every minutes and performing planar-averages in the horizontal. Figure~\ref{fig:cpuvsgpu_verification} compares CPU and GPU output for various  quantities of the model, namely the temperature, the resolved turbulent kinetic energy representative of the turbulence intensity, the liquid water specific humidity showing where the clouds are, the cloud fraction which is fraction of the area with clouds and the net flux for both longwave and shortwave radiation. Comparison shows excellent agreement for both first order and second order statistics (turbulent kinetic energy), with a maximum relative difference of $\simeq$3\% observed locally on the precipitation mean droplet diameter.

\section{Performance}
\label{sec:perf}

\subsection{Single node}

\subsubsection{Hardware and set up} In order to assess the performance of the GPU acceleration for different GPU hardware, we compare model run times obtained on a node-basis on the Dutch national supercomputer Snellius~\cite{snellius}. 
We namely consider the following computing nodes:
\begin{itemize}
    \item a CPU node consisting of 192 AMD Genoa 9654 
    \item a GPU node consisting of 4 NVIDIA A100 
    \item a GPU node consisting of 4 NVIDIA H100 
\end{itemize}
The number of spatial points in the horizontal directions (number of grid columns) is adapted as indicated in Table~\ref{tab:hardware} in order to accommodate with different MPI partition of the spatial domain and to ensure that the GPU memory is filled.

\begin{table*}[!ht]
\centering
\hspace*{-2cm}
\begin{tabular}{{c|c|c|c}}
\centering
{\bf Hardware} & {\bf MPI tasks} & {\bf nb. of grid columns per task} & {\bf total nb. of grid columns}\\
\hline
CPU AMD Genoa (Snellius) & 192 & 63$\times$84 & 1.016 10$^6$\\
GPU NVIDIA A100 (Snellius) & 4 & 512$\times$512 & 1.049 10$^6$\\
GPU NVIDIA H100 (Snellius) & 4 & 512$\times$512 & 1.049 10$^6$\\
\hline
\end{tabular}
\caption{Single-node comparisons. Considered hardware with associated number of MPI tasks, number of grid columns per task and total number of grid columns.}
\label{tab:hardware}
\end{table*}

The performance of DALES on the Botany case are measured during a short simulation lasting 10 min of model time and restarting after the day-long initialization run. Only the time spent in the time stepping loop is included in the measurements and the experiments are repeated 5 times. Because the number of time steps and the number of computational cells differ between the runs (due to the chaotic nature of the flow and the scaling of the computational domain), 
we present timing data in units of $\mu$s per time step per million of cells.

\subsubsection{Timings}

Table~\ref{tab:single-node-timings} shows the averaged performance data for the CPU-only, the A100  and the H100 runs for each component of the DALES solver. For each hardware, the execution time and the fraction of the total time is presented, and the speed up compared to the CPU-only run is also added for the GPU-accelerated cases. The overall speed-up of DALES on a node basis is around 4 on A100 and 11.5 on H100. The fractions of the time spent in each component of the code remaining close to their CPU value with one exception: on A100, the performance of the Poisson solver are below average with a speedup close to 2. Further testing showed that the cuFFT solver available with NVHPC 24.5 (CUDA 12.1.1) on Snellius is less efficient than that available with NVHPC 22.7 (CUDA 11.7.0). The later software stack brings the Poisson solve speedup on par with the remainder of the code, but performance data obtained with the former stack were kept in Table~\ref{tab:single-node-timings} to maintain a consistent software stack across both NVIDIA GPUs. Both stencil-based functions (advection, subgrid) show similar trends with above averaged speedups while thermodynamics  exhibits below averaged speedup, driven by an increased need for synchronisation and horizontal reductions. The significantly higher speed-up obtained on GPUs for the surface component mainly results from the lower number of MPI ranks compared to the CPU case, since the surface computation include reductions in order to obtain integrated ground friction and other quantities of interests.

\begin{table*}[!ht]
\centering
\hspace*{-2cm}
\begin{tabular}{{l|c|c|c|c|c|c|c|c}}
\centering
 & \multicolumn{2} {c|} {\bf CPU} & \multicolumn{3} {c|} {\bf A100GPU} & \multicolumn{3} {c} {\bf H100GPU} \\
\hline
{\bf Timer} &                   Timing &    Fraction  & Timing &    Fraction & speed-up & Timing &    Fraction & speed-up \\
\hline     
Thermodynamics       &       1178.9 &        12.1 &        221.8 &         9.1 &       5.32 &        120.7 &        14.3 &      9.77  \\
Poisson              &       2729.2 &        28.1 &       1442.0 &        59.0 &       1.89 &        220.5 &        26.1 &     12.38  \\
Subgrid              &       1002.3 &        10.3 &        131.6 &         5.4 &       7.62 &         89.9 &        10.6 &     11.15  \\
Radiation            &       1172.3 &        12.1 &        132.8 &         5.4 &       8.83 &         86.1 &        10.2 &     13.61  \\
Microphysics         &        607.2 &         6.3 &         97.0 &         4.0 &       6.26 &         73.5 &         8.7 &      8.26  \\
Advection            &        922.1 &         9.5 &        106.7 &         4.4 &       8.64 &         74.3 &         8.8 &     12.41  \\
Time stepping        &        733.2 &         7.6 &         56.2 &         2.3 &      13.05 &         37.6 &         4.4 &      19.5  \\
ExtForces            &        497.9 &         5.1 &         61.9 &         2.5 &       8.04 &         45.7 &         5.4 &     10.89  \\
Halo-Exchange         &        501.0 &         5.2 &        112.0 &         4.6 &       4.47 &         42.7 &         5.0 &     11.73  \\
Surface              &        246.1 &         2.5 &          3.2 &         0.1 &      76.91 &          4.3 &         0.5 &     57.23  \\
Checks               &         44.5 &         0.5 &          5.9 &         0.2 &       7.54 &          4.6 &         0.5 &      9.67  \\
\hline
Timestep loop        & {\bf 9711.7} & {\bf 100.0} & {\bf 2444.2} & {\bf 100.0} & {\bf 3.97} & {\bf 846.2}  & {\bf 100.0} & {\bf 11.48}\\
\hline
\end{tabular}
\caption{Single node CPU and GPU timings in $\mu$s/step/Mcells. 
}
\label{tab:single-node-timings}
\end{table*}


Timings of the components of the Poisson solver are presented in Table~\ref{tab:poisson_timings}. It can be seen that most of the time is spent in the forward and backward FFT steps which, as mentioned before, require MPI all-to-all communication operations. The forward and backward FFT steps only experience a  small speedup when comparing timings on the A100 GPU node to the CPU node. Speedups on the FFT steps comparable to those observed on H100 GPU node were attained when using the NVHPC 22.7 (CUDA 11.7.0) distribution.

\begin{table*}
    \centering
    \hspace*{-2cm}
    \begin{tabular}{c|c|c|c|c|c|c|c|c}
         & \multicolumn{2}{|c|}{\textbf{CPU}} & \multicolumn{3}{|c|}{\textbf{A100GPU}} & \multicolumn{3}{|c}{\textbf{H100GPU}} \\ \hline
        \textbf{Timer}    & Timing & Fraction & Timing & Fraction & Speedup & Timing & Fraction & Speedup \\ \hline
        Fill RHS          & 516.6  & 18.9     & 42.3   & 2.9      & 12.21   & 23.3   & 10.5     & 22.17   \\
        Forward FFT       & 946.7  & 34.7     & 603.3  & 41.8     & 1.57    & 86.9   & 39.3     & 10.9    \\
        Tridiagonal solve & 110.5  & 4.0      & 10.0   & 0.7      & 11.05   & 6.2    & 2.8      & 17.82   \\
        Backward FFT      & 1025.5 & 37.6     & 659.4  & 45.7     & 1.55    & 89.5   & 40.5     & 11.46   \\
        Apply correction  & 129.9  & 4.8      & 126.6  & 8.8      & 1.03    & 15.1   & 6.8      & 8.60    \\ \hline
        Total             & 2729.2 & 100      & 1441.6 & 100      & 1.20    & 220.9  & 100      & 7.8     \\ \hline
    \end{tabular}
    \caption{Single node CPU and GPU Poisson solver timings in $\mu$s/step/Mcells.}
    \label{tab:poisson_timings}
\end{table*}

Radiation calculation is another computationally expensive step in the time loop (about 12~\% of CPU timings). As explained in Sec.~\ref{sec:acc_librairies}, these calculations are divided into two steps for both longwave and shortwave spectral regions: the computation of radiative properties (RRTMGP) and the radiative transfer equation (RTE). Table~\ref{tab:radiation-timings} gives additional timings for each of these substeps. Total speed-up is about 9 and 14 for A100 GPUs and H100 GPUs, respectively. The speed-up is roughly the same for each of the individual substep, although RRTMGP LW and RTE SW are proportionately a little faster. This homogeneous speed-up is expected since radiation calculations are independent for each vertical column and each spectral band.


\begin{table*}[!ht]
\centering
\hspace*{-2cm}
\begin{tabular}{{l|c|c|c|c|c|c|c|c}}
\centering
 & \multicolumn{2} {c|} {\bf CPU} & \multicolumn{3} {c|} {\bf A100GPU} & \multicolumn{3} {c} {\bf H100GPU} \\
\hline
{\bf Timer} &                   Timing &    Fraction  & Timing &    Fraction & speed-up & Timing &    Fraction & speed-up \\
\hline 
RRTMGP LW & 332,2 & 28.3 & 35.3 & 26.6 & 9.41 & 22.2 & 25.8 & 14.96 \\
RRTMGP SW & 277.0 &23.6 & 34.0 & 25.6 & 8.15 & 21.4 & 24.9 & 12.94 \\
RTE LW & 166.7 & 14.2 & 19.6 &  14.8 & 8.51 & 13.4 & 15.6 & 12.44 \\
RTE SW & 345.6 & 29.5 & 37.7 &  28.4 & 9.17 & 25.0 & 29.0 & 13.82 \\
\hline
Radiation total & 1172.3 & 100.0 & 132.8 & 100.0 & 8.83 & 86.1 & 100.0 & 13.62 \\
\hline
\end{tabular}
\caption{Single node CPU and GPU radiation timings in $\mu$s/step/Mcells.}
\label{tab:radiation-timings}
\end{table*}

Finally, note that while the size of the DALES test case has been kept constant on all hardware, more GPU memory is available on H100 
and larger test cases could be run on these hardware, further increasing the speedup using the performance metric listed in Table~\ref{tab:single-node-timings}.

\subsection{Weak scaling}

\begin{figure}[!ht]
    \centering
    \includegraphics[width=0.95\linewidth]{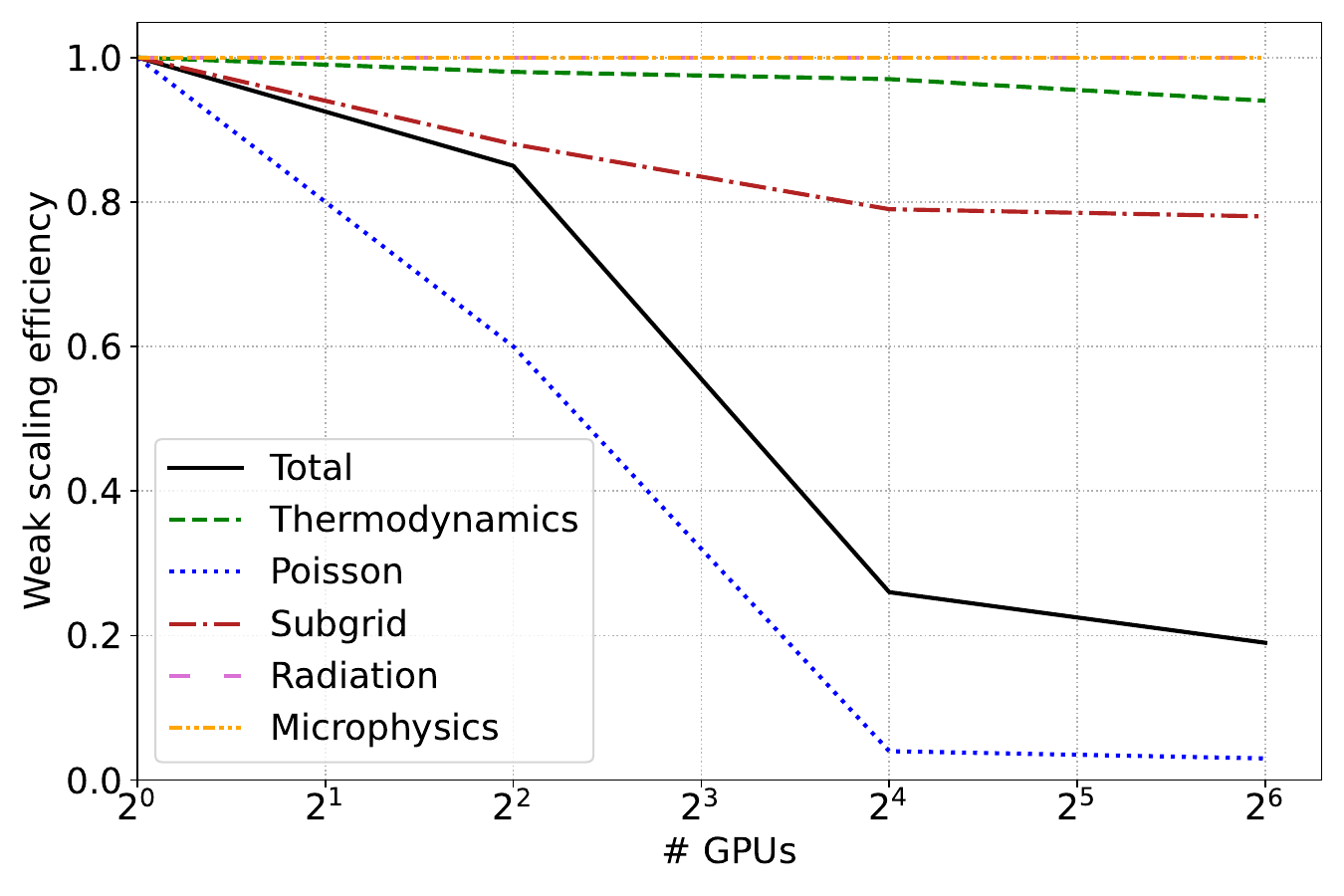}
    \caption{Weak scaling of DALES on the botany case using Snellius A100 GPUs}
    \label{fig:weak_scaling}
\end{figure}

To evaluate the ability of DALES to scale with the needs of researchers, the A100 case is scaled from 1 to 64 GPUs (0.25 to 16 compute nodes), leveraging the periodicity of the  test case to duplicate the initial data in the horizontal directions in order to keep the workload per GPU constant. Figure \ref{fig:weak_scaling} shows the results of the overall DALES weak scaling efficiency ($t_{1_{GPU}}/t_{N_{GPU}}$) as well as the efficiency of individual components. DALES parallel efficiency is found to drop quickly, mostly driven by the poor scaling performance of the Poisson solve, whereas the other components maintain an efficiency close to one as very little to no communications are needed. Improvement of DALES scaling performance, in particular exploring alternate projection approaches or improving the communication pattern of FFT-based solvers \cite{Romero2022}, is left for future work.

It is worth mentioning here that strong scaling is disregarded in this study, since the objective of acceleration is to address larger spatial domains.

\section{Optimization}
\label{sec:opt}

The acceleration strategy presented in Sec.~\ref{ssec:accel} enables to quickly accelerate the entire code base, giving entire freedom to the compiler in terms of kernel launch parameters and ensuring relatively good portability. However, a handful of kernels exhibit low occupancy ($<$50~\%) and require more careful consideration. Two strategies are employed in order to increase the performance of those kernels: 1) manually modifying the kernels to improve memory access, cache usage and register pressure; 2) using Kernel Tuner \cite{vanwerkhoven19} (KT) to auto-tune the kernel launch parameters.

\subsection{Presentation of Kernel Tuner}
Kernel Tuner (KT)~\cite{vanwerkhoven19} is a generic auto-tuning framework for GPU applications.
It is written in Python and supports the tuning of computational kernels written in CUDA, HIP, and OpenCL, but was also recently extended to handle directive-based languages like OpenACC and OpenMP.

To do so, KT provides a set of helper functions that allows users to extract the tunable kernels from the rest of the code, together with all the necessary C preprocessor macros, and takes care of the allocation and initialization of the used data structures.

The selection of tunable kernels and identification of the data structures is guided by the user, by adding some ad-hoc KT pragmas to the source code; these pragmas are the only changes to the original code needed for tuning.

\hspace*{-2cm}
\begin{lstlisting}[float=*, language=Fortran, caption=Momentum diffusion operator kernel where $km$ and $e$ are respectively the eddy viscosity and the subgrid TKE, label=lst:OptimKernel]
!$acc parallel loop num_gangs(ngang) vector_length(vlength) collapse(cfactor)
do k = 1, kmax
  do j = 1, jmax
    do i = 1, imax
       e_t(i,j,k) =(((km(i+1,j,k)+km(i,j,k))*(e(i+1,j,k)-e(i,j,k)) &
                    -(km(i,j,k)+km(i-1,j,k))*(e(i,j,k)-e(i-1,j,k))) &
                    * nuf(k)/dx2 &
                  + ((km(i,j+1,k)+km(i,j,k))*(e(i,j+1,k)-e(i,j,k)) &
                    -(km(i,j,k)+km(i,j-1,k))*(e(i,j,k)-e(i,j-1,k))) &
                    * nuf(k)/dy2 + &
                  + ( rhoh(k+1)/rhof(k) * (dzf(k+1)*km(i,j,k)+dzf(k)*km(i,j,k+1) &
                                        * (e(i,j,k+1)-e(i,j,k))) / dzh(k+1)**2 &
                     -rhoh(k)/rhof(k)   * (dzf(k-1)*km(i,j,k)+dzf(k)*km(i,j,k-1) &
                                        * (e(i,j,k)-e(i,j,k-1))) / dzh(k)**2 &
                    ) &
                   )
    enddo
  enddo
enddo
\end{lstlisting}

Kernel Tuner was used to find the optimal configuration of launch parameters for the GPUs used in the experiments. All the KT data reported hereafter are averaged over 20 calls to the targeted kernel with a given set of parameters in order to mitigate runtime variability.

\subsection{Targeted kernels}
\label{ssec:target_KT}
The compute kernels targeted for optimization are the stencil-based kernels in the advection and subgrid scale diffusion operator. Combined, these kernels take up to 20\% of the total compute time (see Table~\ref{tab:single-node-timings}). Note that tuning of launch parameters in typical thermodynamics and microphysics kernels was also attempted without obtaining any gain compared to the compiler-selected values.

The subgrid diffusion operator is close to a weighted 3D 7-points stencil, which is a benchmark for a number of stencil kernel domain specific language \cite{Yount2016,Zhao2018}. These studies showed that the 3D 7-points stencil is difficult to optimize because only close neighbors (+/-1 in each direction) are involved, such that data locality is not as critical as higher order kernels. Additionally, these kernels are memory bound and thus less sensitive to kernel launch parameters. For instance the subgrid turbulent kinetic energy (TKE) production through strain (second term on the right-hand-side of Eq. 12 in~\cite{heus10}) kernel is given in Listing~\ref{lst:OptimKernel}.

In this kernel, the stencil in the $z$ direction differs from the horizontal ones due to the staggered nature of the grid and the non-uniform grid spacing. A variant of the kernel in Listing \ref{lst:OptimKernel} for computing subgrid scalar diffusion adds a fourth loop on the number of scalars. This fourth loop can be placed around the $ijk$ \verb|do| loops, further extending the index space when using \verb|collapse|, or nested as an inner loop and treated sequentially using the \verb|loop seq| clause such that each GPU thread performs the computation for all the scalars. The advection kernels (not listed here for the sake on concision) features similar access pattern as those presented in Listing \ref{lst:OptimKernel}, but the higher order flux reconstruction (6$^{th}$-order) needs a wider stencil, reaching +/-3 in each direction.

An alternative approach to dealing with stencil operation consists in replacing the \verb|collaspe| clause with a \verb|tile(tx,ty,tz)|, directing the compiler to split each of the nested loop into a outer tile loop and an inner element loop, in hope that it might result in better cache reuse when already present neighboring cell data is accessed again. The same objective can also be targeted by manually tilling the inner $ij$ loops with $nbi$-$nbj$ blocks of size $i_b$-$j_b$ as depicted in Listing~\ref{lst:manualTile}.

\begin{lstlisting}[float, floatplacement=!H, language=Fortran, caption=Example of manual loop tiling, label=lst:manualTile]
!$acc parallel loop num_gangs(ngang) &
!$acc& vector_length(vlength) &
!$acc& collapse(cfactor)
do k = 1, kmax
  do bj = 0, nbj
    do bi = 0, nbi
      !$acc loop seq collapse(2)
      do j = 1, j_b
        do i = 1, i_b
          ...
        enddo
      enddo
    enddo
  enddo
enddo
\end{lstlisting}

Here, a \verb|loop seq| clause is used on the inner $ij$ element loops such that each GPU thread performs the computation for all the elements in the tile.

In summary, the kernel launch parameters targeted by KT are:
\begin{itemize}
    \item \verb|num_gang| maps to the length of the thread block grid in the $x$ direction (\verb|gridDim.x| with a CUDA backend)
    \item \verb|vector_length| maps to the block size in the $x$ direction (\verb|blockDim.x| with a CUDA backend)
    \item \verb|collapse| indicates how many of the tightly nested loop index spaces are flatten to a single one
    \item \verb|tx,ty,tz| the tile size in the three spatial dimension employed by OpenACC
\end{itemize}

\subsection{Experiments on selected kernels}

\begin{figure}[!ht]
    \centering
    \includegraphics[width=0.98\linewidth]{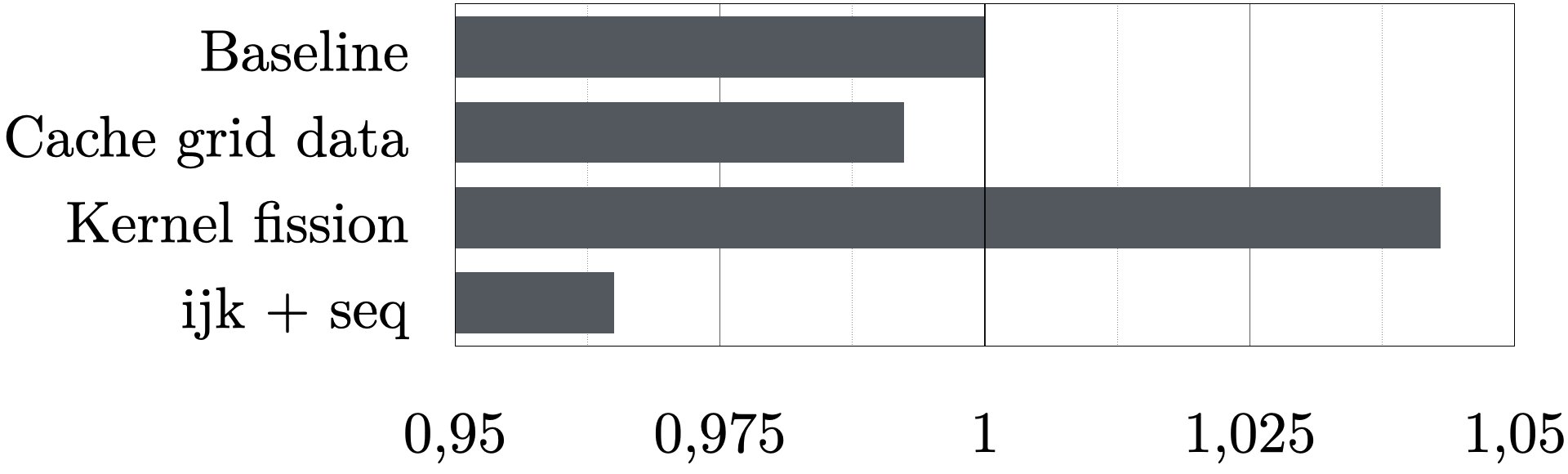}
    \caption{Run times of the subgrid module with three optimisation strategies (using the cache directive, using kernel fission and using \emph{ijk} collapsed loop and \emph{n} sequential loop), relative to the baseline acceleration strategy.}
    \label{fig:sgs_optim_manual}
\end{figure}

In a first step, the scalar subgrid diffusion kernel \verb|diffcsv| is manually tweaked in order to improve performance. The following changes are implemented: 1) using the \verb|cache| directive to load anisotropic grid spacing data in software-managed caches instead of multiple read into 1D array, 2) kernel fission of the subgrid kinetic energy sources to reduce register pressure (from 108 down to 62 + 56), 3) replace \emph{ijkn} collapsed nested loop with a \emph{ijk} collapsed one and a \verb|loop seq| on the $n$ component. Figure~\ref{fig:sgs_optim_manual} shows the results. Using a combination of 1) and 3), the overall performance of the subgrid module can be increased by close to 5\%. Reducing register pressure by splitting the long kernel into two leads to lower performance as the resulting kernels are effectively run sequentially since each can still occupy the entire GPU.

We then use KT on two specific subgrid kernels: \verb|diffcsv| (which is similar to Listing \ref{lst:OptimKernel} but including an outer loop on scalar and 4-dimensional Fortran arrays) and \verb|sources|, their contribution accounting for 30\% of the subgrid module. The first kernel is of interest as it evaluates the diffusion term for transported scalars in DALES, the number of which can vary from 2 (as in the Botany case) to a few dozens when chemical species are considered; the second kernel exhibit significant reuse of the neighboring data to evaluate cross-derivatives and is a good candidate for improved performance using tiling. All the tests are conducted on a grid size similar to that of the botany case on a single GPU, and using both A100 and H100 GPUs in order to evaluate the portability of the tuned kernels.

\begin{figure}[!ht]
    \centering
    \includegraphics[width=0.98\linewidth]{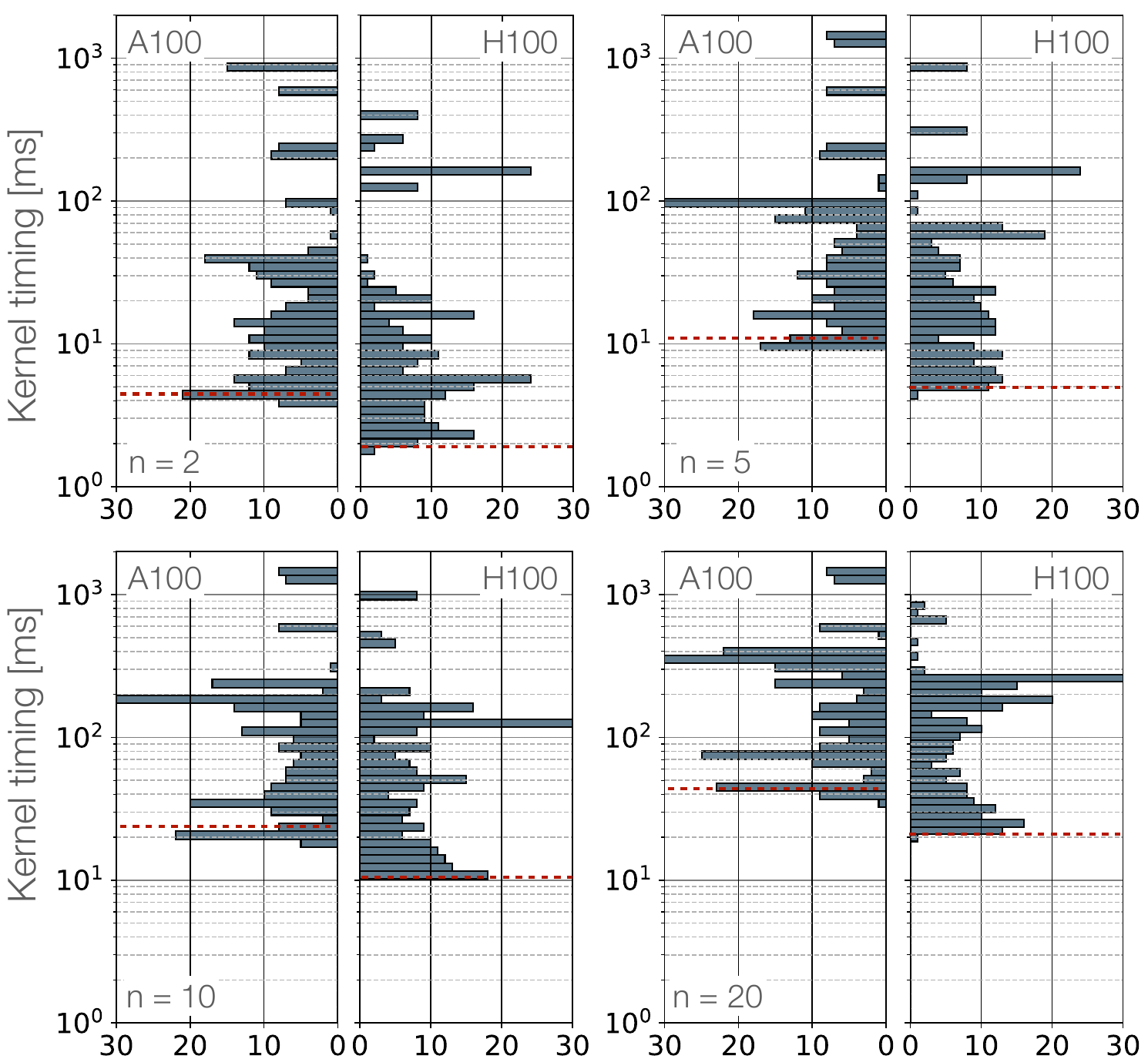}
    \caption{Measured kernel timing distribution for the diffcsv kernel, varying the number of scalar $n$. On each subplot, the left and right panel corresponds to A100 and H100 GPUs respectively. Dashed red lines indicate default parameters timing.}
    \label{fig:sgs_optim_KT}
\end{figure}

Figure \ref{fig:sgs_optim_KT} shows the distribution of the \verb|diffcsv| kernel timings for an increasing number of transported scalars and kernel launch parameters in the following range: \verb|vector_length| $\in [16:512]$, \verb|num_gangs| $\in [256:65535]$, \verb|cfactor| $\in [2:4]$ and with or without the \verb|loop seq| optimization shown in Fig.~\ref{fig:sgs_optim_manual} (a total of 256 parameter sets per subplot). Each subplot is a classical histogram plot, oriented vertically such that the bar length (abscissa) is the number of parameter sets that leads to a timing within a given time bin (ordinate). The left panel corresponds to A100 timings, while the mirrored right one is H100 timings. Additionally the timing with the default (compiler-selected) parameters are indicated with a dashed red line. Optimizing the kernel launch parameters only resulted in significant performance improvements on A100, with a maximum gain close to 35\% in the $n=10$ case. On H100, only small improvements were obtained, up to 10\%. Using \verb|loop seq| always led to better performance, especially for a small number of scalars. Interestingly, the timing distribution on H100 are not simply a translation of the A100 one, and the optimum parameter set differs slightly between the two GPUs: larger \verb|vector_length| and \verb|num_gangs| led to optimum on H100 compared to A100. However, the parameters leading to optimum performance on H100 gave close to optimum performance on A100 (within 5\%), and vice-versa.

\begin{figure}[!ht]
    \centering
    \includegraphics[width=0.98\linewidth]{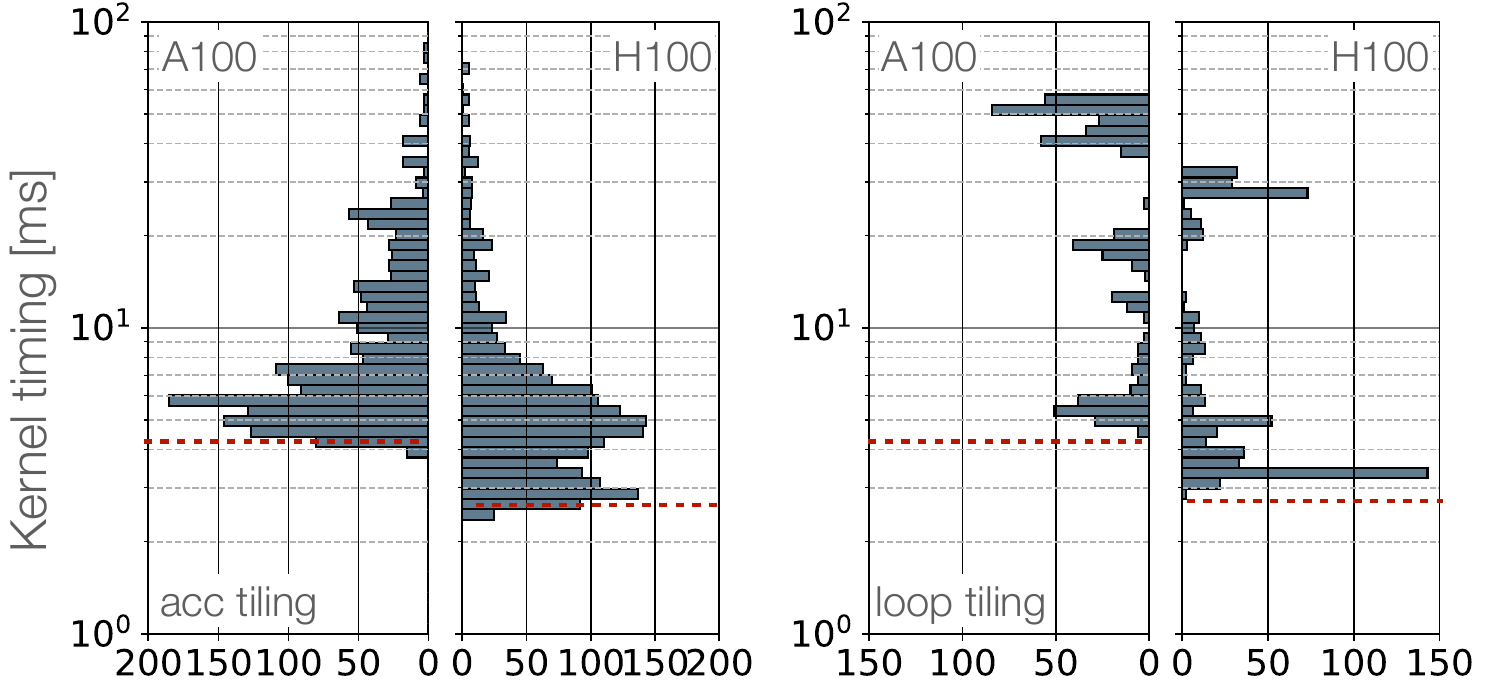}
    \caption{Measured kernel timing distribution for the sources kernel, using OpenACC tiling clause (left) and introducing $ij$ tiles (right). On each subplot, the left and right panel corresponds to A100 and H100 GPUs respectively. Dashed red lines indicate default parameters timing.}
    \label{fig:sgs_optim_KT_sources}
\end{figure}

On the second kernel, we could not find launch parameters leading to better performance than the defaults with KT when using only the \verb|collapse| clause. Instead, we employed 1) the \verb|tile(tx,ty,tz)| clause discussed in Section \ref{ssec:target_KT} and 2) the manual tilling of Listing \ref{lst:manualTile}. Figure \ref{fig:sgs_optim_KT_sources} shows the kernel timing for optimization 1) and 2) on the left and right, respectively. KT allows to automatically test hundreds of combinations of \verb|(tx,ty,tz)| and $i_b$, $j_b$ values, for values of \verb|vector_length| and \verb|num_gangs| in the same range as for Fig.~\ref{fig:sgs_optim_KT}. Using OpenACC \verb|tile| led to up to 15\% performance gain on A100 and H100 compared to the baseline \verb|collapse(3)| clause with default vector length and number of gangs. On both GPUs, the best performance are obtained using $tx=64$, $ty=4$ and $tz=2$, and only differs on the value of \verb|vector_length|. Once again, the optimum values of \verb|vector_length| obtained on A100 led to performance on H100 close to the optimum, and the reciprocal. In contrast, manually introducing tiles led to a small decrease in performance at best and much worse performance in most cases.

\begin{figure}[!ht]
    \centering
    \includegraphics[width=0.98\linewidth]{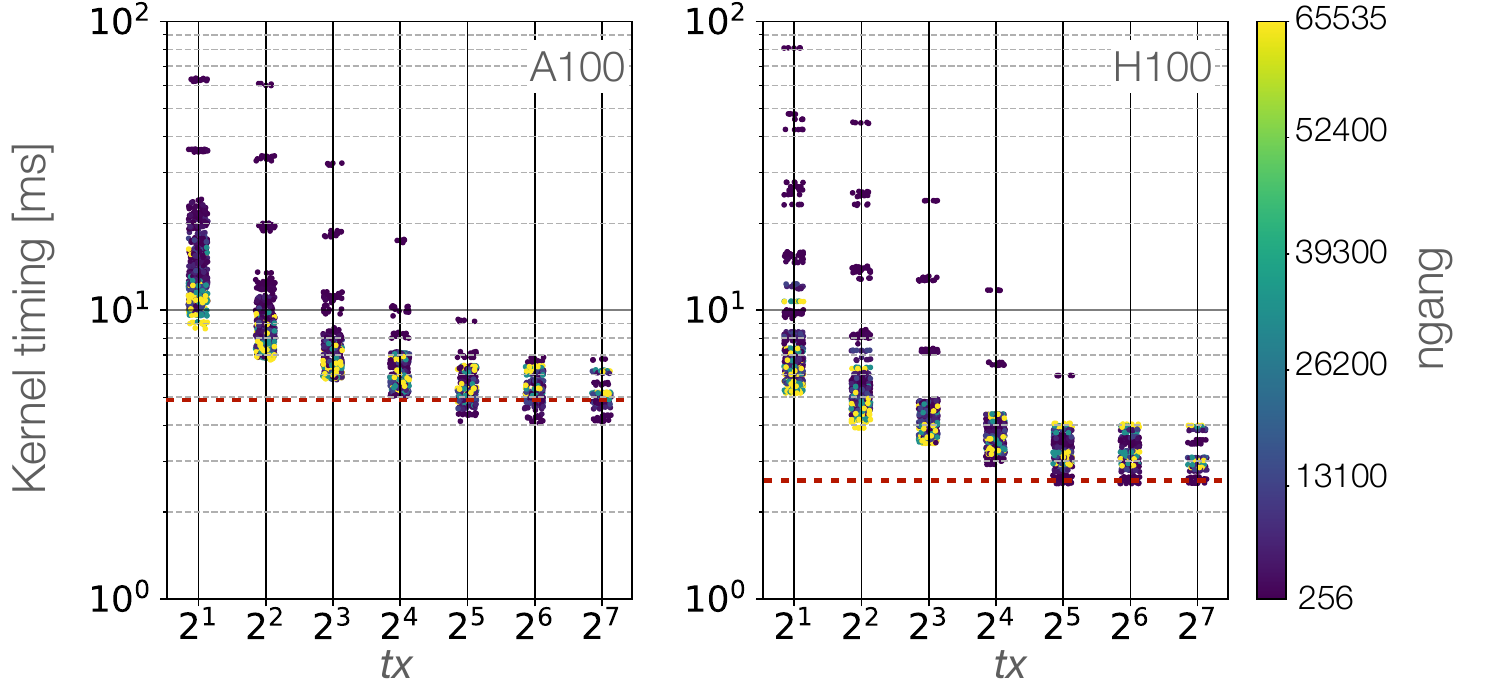}
    \caption{Advection kernel timing for the (512 x 512) $ij$ loops span, as function of the tile size in $x$ direction, colored by the number of gangs used in the launch parameters, left: A100, right: H100. Dashed red lines indicate default parameters timing.}
    \label{fig:adv_optim_KT}
\end{figure}

The \verb|tile(tx,ty,tz)| clause is then employed on another stencil-based kernels,  the 6$^{th}$-order, wider stencil, advection kernel. Additionally, we evaluate if the problem size can affect the potential performance gain by dividing the initial $ij$ loop span (512 x 512) by 2 and 4 in each direction. The results obtained on the advection kernel are qualitatively similar with those obtained on the subgrid \verb|source| kernel: up to 15\% performance gain on A100 and only marginal gain on H100 GPUs. To illustrate the effect of the tile size on performance, Fig. \ref{fig:adv_optim_KT} shows the kernel timing for the (512 x 512) loop span as a function of the tile size $tx$ which was found to correlate best with performance. For each value of $tx$, a small random perturbation was added to visualize the timing dispersion due to the other tuning parameters. In particular, data are colored by the number of gangs. Large values of $tx$ significantly improve the kernel performance, while a large number of gangs is beneficial at small values of $tx$. However, a smaller number of gangs leads to better performance when large $tx$ are used. The KT timing results for smaller loops span are qualitatively similar to Fig. \ref{fig:adv_optim_KT}, but no performance gain could be obtained compared to the default kernel launch parameters.

Finally, we implemented the set of parameters that led to performance improvements throughout the code base in order to evaluate an overall gain from the optimization with KT. More specifically, we introduced a \verb|collapse(3)| with a \verb|loop seq| in the scalar space in place of all the \verb|collapse(4)| for the $ijkn$ loops, and we replaced \verb|collapse(3)| with a \verb|tile(tx,ty,tz)| clause for all the stencil-based kernels, keeping the A100 best launch parameters. The resulting overall speed-up of the Botany case listed in Table \ref{tab:single-node-timings} was 1.7\% on A100 and 2.6\% on H100. A closer look at specific kernel timings shows that propagating the kernel launch parameters optimized with KT on the handful of kernels reported here to all of the stencil-based kernels has a mixed outcome. Several kernels are effectively faster, close to the performance gains observed with KT, but the performance of some other kernels (also stencil-based kernels, but with different access pattern and intermediate computations) is degraded. This last point further highlights the difficulty to perform optimization at the application level and point towards an avenue of improvement for auto-tuning tools such as KT.

\section{Conclusion}
\label{sec:conclu}

This paper presented the GPU porting of a large-eddy-simulation atmospheric model (DALES), which is a multi-physics Fortran code designed to run on a large spatial domain with spatial resolution typically of the order of 5-100~m and for a long integration time of the order of several days. The use of OpenACC directives facilitated the porting, mainly consisting of OpenACC collapsed loops over the three dimensions of space and fusing of as many successive, independent kernels as possible. The porting also benefited from GPU-aware libraries for radiation calculations (RTE-RRTMGP) and Fourier transforms (cuFFT). Still, refactoring some parts of the code was necessary, especially in the microphysics and thermodynamics sections, where differences in the local state introduced significant thread divergence.

Acceleration performance was assessed from a test case featuring cloud transport and radiative transfer. Run times between CPU and GPU implementations were compared on a single-node basis. A speed-up of 4 and 12 has been obtained for nodes of 4 NVIDIA A100 and 4 NVIDIA H100 respectively, compared to a node of 192 AMD CPUs. The most computationally intensive sections of the code were found to be the thermodynamics, the subgrid diffusion, the radiative transfer and the Poisson solver. Overall, the fractions of the time spent in each component of the code remained close to their CPU value. However, the Poisson solver was found to perform poorly on the A100 because of a loss of efficiency of the cuFFT library in the associated software stack. Thermodynamics routines also exhibited below-average speed-up because of required synchronisations and horizontal reductions.

The weak scaling analysis showed that the acceleration efficiency drops quickly beyond a single node. The Poisson solver, based on Fourier transforms, is responsible for this loss of efficiency, as performing FFT across the whole spatial domain requires all-to-all communications and memory rearrangements, which do not scale. We might consider implementing alternative Poisson solvers that could better scale with many GPUs. Having said that, we stress that the single-node acceleration remains highly valuable for research, as the simulations must be performed for a large integration time (up to several hundred hours) to accumulate statistics.

A noteworthy aspect of this work is the use of Kernel Tuner, an auto tuning application, to optimize the GPU execution of stencil-based kernels associated with subgrid diffusion and advection processes.
For the subgrid diffusion kernel, significant performance improvements were obtained on A100 GPUs, with a maximum gain close to 35~\%, while limited performance improvements were obtained on H100 GPUs, with a maximum gain close to 10~\%. We noticed that the optimal parameters differ slightly between the two GPUs, even though using the optimal parameters of one GPU on the other one resulted in near optimal performance. 
For subgrid diffusion source and 6$^{th}$ order advection kernels, we attempted to replace collapsing clauses with tile clauses and it only resulted in better performance on the A100 GPUs, around 15~\%. 

Finally, we examined how the tuned parameters can be transferred to similar kernels (i.e. stencil based kernels) across the entire DALES code base. The overall DALES performances gain only showed small improvements compared to the initial port, and fine grain profiling revealed the mitigated effects of the tuned parameters. These observations deserve further investigation and underline the challenge associated with application level performance tuning.

Ongoing efforts in DALES focus on evaluating and improving the performances of the present OpenACC port on AMD GPU hardware, as well as  investigating alternate strategies for solving the Poisson linear system in order to improve DALES scaling. In future steps, we would also like to investigate the effect of mixed precision on acceleration performance and code accuracy. DALES already allows switching some quantities to single-precision, which will result in the ability to address larger resolutions on the GPU.

\section{Acknowledgements}

The authors acknowledge funding by the European Union. This work has received funding from the European High Performance Computing Joint Undertaking (JU) under grant agreement No 101093054.

The authors acknowledge the support of the Ruisdael Observatory, 
scientific research infrastructure which is (partly) financed by
the Dutch Research Council (NWO, grant no. 184.034.015)

This work used the Dutch national e-infrastructure with the support of 
the SURF Cooperative using grants no. vus21036 and EINF-10608.

Caspar Jungbacker acknowledges support by the Refreeze the Arctic Foundation (RAF) \url{https://refreezethearcticfoundation.com/about/}.



\bibliographystyle{plain}
\bibliography{biblio}

\end{document}